\documentclass[aps,prb,superscriptaddress,amsfonts,amsmath,amssymb,twocolumn,showpacs,floatfix]{revtex4}
\usepackage{url}
\usepackage{bm}
\usepackage{graphicx}
\usepackage{amsmath}
\usepackage{amstext}
\usepackage{amssymb}
\usepackage{amsfonts}
\usepackage{amsbsy}
\usepackage{verbatim}
\usepackage{color}
\usepackage[colorlinks=true, urlcolor=blue, linkcolor=blue, citecolor=blue, pdftex]{hyperref}
\usepackage{multirow}

\include{diagrammatics}

\begin{document}

\title{Simplex $\mathbb{Z}_2$ spin liquids on the Kagome lattice with Projected Entangled Pair States:
spinon and vison coherence lengths, topological entropy and gapless edge modes}

\author{Didier \surname{Poilblanc} }
\affiliation{Laboratoire de Physique Th\'eorique, CNRS, UMR 5152 and Universit\'e de Toulouse, UPS,
F-31062 Toulouse, France}

\author{Norbert \surname{Schuch}}
\affiliation{Institut f\"ur Quanteninformation, RWTH Aachen, D-52056 Aachen, Germany}



\date{\today}

\begin{abstract} 
Gapped $\mathbb{Z}_2$ spin liquids have been proposed as candidates for the ground-state of the $S=1/2$ quantum antiferromagnet on the Kagome lattice.  
We extend the use of Projected Entangled Pair States to construct (on the cylinder)
Resonating Valence Bond (RVB) states including both nearest-neighbor and next-nearest neighbor singlet bonds. 
Our ansatz -- dubbed "simplex spin liquid" -- allows for an asymmetry between the two types of triangles (of order $2-3\%$ in the energy density after optimization) leading to the breaking of inversion symmetry.
We show that the topological $\mathbb{Z}_2$ structure is still preserved
and, by considering the presence or the absence of spinon and vison lines along an infinite cylinder, we explicitly construct four orthogonal RVB Minimally Entangled States.
The spinon and vison coherence lengths are extracted from a finite size scaling 
w.r.t the cylinder perimeter of the energy splittings of the four sectors   
and are found to be of the order of the lattice spacing. 
The entanglement spectrum of a partitioned (infinite) cylinder is found to be gapless suggesting the 
occurrence, on a cylinder with {\it real} open boundaries, of gapless edge modes formally similar to Luttinger 
liquid (non-chiral) spin and charge modes. 
When inversion symmetry is spontaneously broken, the RVB spin liquid exhibits an extra Ising degeneracy, 
which might have been observed in recent exact diagonalisation studies. 

\end{abstract}
\maketitle

\begin{figure}
\includegraphics[width=0.95\columnwidth]{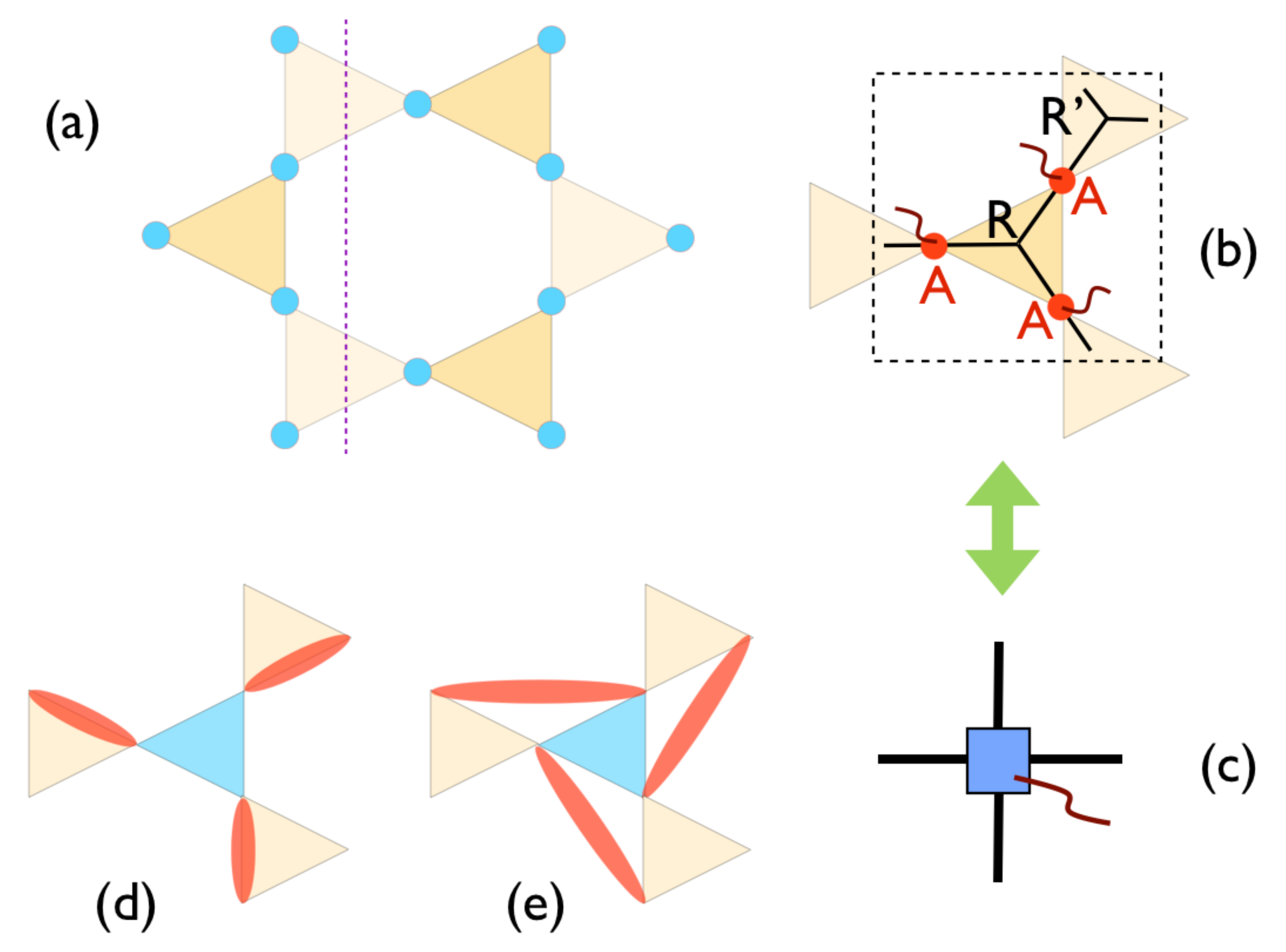}
\caption{(a) Kagome lattice with a vertical cut (dashed line).  If the inversion center is lacking, right (light shading) and left (darker shading) triangles become non-equivalent but all sites (blue dots) keep the same environment. (b,c) PEPS representation of the RVB state: three $S=1/2$ sites are grouped together (b) to construct a rank-5 tensor (c) involving $2^3=8$ physical states
and $D=3$ virtual states on the four out-going bonds. (d,e) Two dimer configurations around a "defect  triangle" 
(in blue). Red ellipses represent singlets between two sites. 
The $\alpha$ parameter (see text) controls the relative weights between (d) and (e).}
\label{Fig:peps}
\end{figure}

{\it Introduction --}
Gapped spin liquids are intriguing interacting quantum spin states which do not bear any local order parameter but possess {\it topological
order}~\cite{wen-91}. The great excitement for such systems is partly due to their potential to realize 
qubits for future topological computers~\cite{kitaev:toriccode}. Frustrated quantum magnets~\cite{review} like the spin-1/2 
Quantum Antiferromagnetic Heisenberg
(QAH) model on the kagome lattice (see Fig.~\ref{Fig:peps}(a)) are believed to 
be good model candidates to host topological spin liquids. The Anderson's Resonating Valence Bond (RVB) state~\cite{anderson},
Gutzwiller projected BCS wavefunctions~\cite{z2-vmc} or the solvable dimer-liquid~\cite{RK} provide simple 
realizations of such states. 
Nevertheless, Lanczos exact diagonalisations (LED) of small clusters~\cite{review} of the  kagome QAH model have 
revealed a large amount of low-energy singlets,
a strong signal for other competing non-magnetic ground-states (GS), like the gapless U(1) spin liquid~\cite{U1-lee} or 
valence bond crystals (VBC) spontaneously breaking lattice translation~\cite{vbc-series,vbc-class,vbc-qdm}. 
However, recent Density Matrix Renormalization Group (DMRG) studies~\cite{White,Balents,Schollwoeck} 
have provided more evidence in favor of a
topological gapped $\mathbb{Z}_2$ spin liquid, in the neighborhood of a U(1)-Dirac spin liquid state~\cite{Lu_2011} 
nearby in energy~\cite{U1-iqbal}.

In bisimplex lattices, like kagome or pyrochlore lattices, consisting of corner-sharing simplices (e.g. triangles or tetraedra) located on an
underlying bipartite lattice, a spontaneous symmetry breaking can be
expected,~\cite{indergand06} resulting in a phase where the ÒleftÓ-simplices
differ from the ÒrightÓ-simplices (see Fig.~\ref{Fig:peps}(a)). For the kagome lattice the inversion symmetry is broken
in this phase. Interestingly, state-of-the-art LED of kagome tori with up to 48 sites~\cite{laeuchli} have suggested 
that the GS of the kagome QAH model might be a gapped spin liquid with 8-fold GS degeneracy originating
from the product of the usual 4-fold topological $\mathbb{Z}_2$ degeneracy (on the torus) by a 2-fold Ising 
degeneracy associated to spontaneous breaking of inversion symmetry. 
Following this proposal, we construct here a simple ansatz for a topological $\mathbb{Z}_2$ spin-liquid lacking inversion symmetry
that might reflect qualitatively the properties of the GS of the QAH model on the kagome lattice.
For this purpose we use the Projected Entangled Pair States (PEPS) formalism~\cite{PEPS} which provides
a natural construction of RVB wavefunctions~\cite{verstraetewolf06,long_paper} as well as a simple understanding of 
their symmetries~\cite{schuch:peps-sym}, topological and entanglement properties~\cite{long_paper_topo}, and boundary theories~\cite{ciracpoilblanc2011,schuch2012,long_paper,long_paper_topo}.

{\it PEPS construction --} 
We start with the  RVB wavefunction defined on the kagome lattice of Fig.~\ref{Fig:peps}(a) as an equal weight (and equal sign) summation of 
all nearest-neighbor (NN) singlet (\hbox{$|\uparrow\downarrow\rangle-|\downarrow\uparrow\rangle$}) coverings (NN  
singlets are all oriented clockwise on all the triangles).  Such a state can in fact be
represented by a $D=3$ PEPS~\cite{verstraetewolf06,long_paper} (up to local unitaries) in terms of 
two rank-3 tensors, (i) $A^{s}_{\lambda,\mu}$ on the sites, and (ii) $R_{\lambda,\mu,\nu}$
in the center of the left and right triangles, where $s=0,1$ are  
qubits representing the two $S_z=\pm 1/2$ spin components and $\lambda,\mu,\nu\in \{0,1,2\}$
are virtual indices (see Fig.~\ref{Fig:peps}(b)). More precisely, $A^{s}_{2,s}=A^{s}_{s,2}=1$, and zero otherwise, and $R_{2,2,2}=1$,
and $R_{\lambda,\mu,\nu}=\epsilon_{\lambda,\mu,\nu}$ otherwise, 
with $\epsilon_{\lambda,\mu,\nu}$ the antisymmetric tensor.~\cite{long_paper} 
One can then group 3 sites 
on each unit cell to obtain a rank-5 tensor (the physical dimension is now $2^3=8$) connected on an effective square lattice (see Fig.~\ref{Fig:peps}(c)). 
Note that one can arbitrarily choose to group the 3 sites on the left or on the right triangles.
The amplitudes of the RVB state in the local $S_z$-basis are then
obtained by contracting all virtual indices (except the ones at the boundary of a finite system).
This RVB state studied in details in Ref.~\onlinecite{long_paper_topo} is perfectly isotropic i.e. the 
energy densities are identical on all bonds (if the Hamiltonian itself is isotropic).
It is also equivalent to a projected BCS wavefunction~\cite{YangYao}. 

In the case of the NN Heisenberg S=1/2 antiferromagnet on the kagome lattice, "defect triangles" of
Fig.~\ref{Fig:peps}(d) are energetically costly. 
The above RVB wave function has a fixed proportion (1/4) of defect triangles 
characterized by $\lambda=\mu=\nu=2$ on the 3 bonds of the corresponding $R$-tensor.
However, as shown by Zeng and Elser~\cite{ZengElser}, the variational energy can
be drastically lowered by allowing local fluctuations around each defect triangle, involving 
next-nearest neighbor (NNN) singlets as shown in Fig.~\ref{Fig:peps}(e). Such an improvement can be performed easily within the PEPS formalism 
by acting with the operator $\mathbb{I}-\alpha \mathbb{P}_{3/2}$
on every left triangle (see Fig.~\ref{Fig:peps}(b)), where $\mathbb{I}$ is the identity operator, $\mathbb{P}_{3/2}$ 
is the projector on the fully symmetric subspace of three spins 1/2 
and $\alpha$ is a variational parameter. 
Note that $\mathbb{I}-\alpha \mathbb{P}_{3/2}$ can be re-written as $(1-\frac{\alpha}{2})\mathbb{I}-\frac{2\alpha}{3}\mathbb{H}_\triangleleft$,
where $\mathbb{H}_\triangleleft$ is the local Hamiltonian of the left triangle, and does not affect triangles with a dimer. 
Further optimization can be performed by monitoring the respective proportion of 
``dressed" left and (undressed) right defect triangles, using a different $R'$ tensor on the right triangles, $R'_{2,2,2}=1-\beta$
($0\le\beta\le 1$) and $R'_{\lambda,\mu,\nu}=R_{\lambda,\mu,\nu}$ otherwise. 

\begin{figure}
\includegraphics[width=0.95\columnwidth]{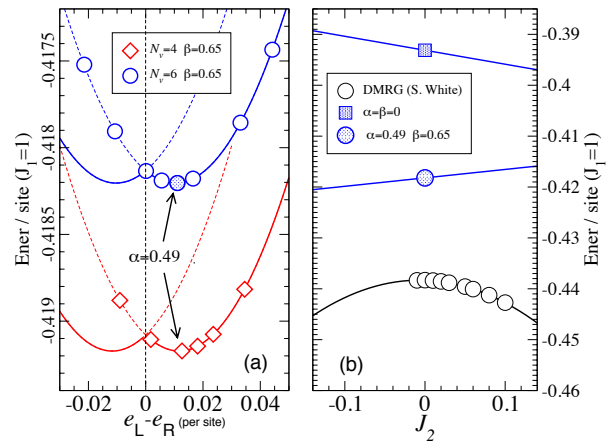}
\caption{(a) Variational energy vs energy difference in left and right triangles 
(varying $\alpha$, at fixed $\beta=\beta^*$).  
The two parabolic "branches" are obtained by interchanging the role of left and right triangles in the PEPS
construction giving two orthogonal wavefunctions $|\Phi_{\rm L}\big>$ and $|\Phi_{\rm R}\big>$. (b) Energy of the
(fixed) optimized state vs NNN AF coupling $J_2$ ($N_v=6$, YC12) compared to DMRG data (S.~White).}
\label{Fig:ener}
\end{figure}

{\it Energetics and inversion symmetry breaking--} The variational energy of the modified PEPS RVB wave function can be optimized 
w.r.t. $\alpha$ and $\beta$ considering infinitely-long YC$n$ ($n=2N_v$) cylinders, i.e. horizontal cylinders w.r.t. lattice orientation
of Fig.~\ref{Fig:peps}(a) as in DMRG studies~\cite{White}.
Perimeters with $N_v=4$ (YC$8$), $N_v=6$ (YC$12$) and $N_v=8$ (YC$16$) unit cells are considered. 
One finds $\alpha=\alpha^*\simeq 0.49$ and 
$\beta=\beta^*\simeq 0.65$ (finite size effects are negligible). Although, by construction, left and right triangles
are non-equivalent, the anisotropy remains quite small at the optimized point.
The variational energy (per site, in units of the coupling $J=1$) for fixed $\beta=\beta^*$ shown in Fig.~\ref{Fig:ener}(a)
as a function of the difference $\Delta e=e_{\rm R}-e_{\rm L}$ between the energy (per site) of the left and right triangles, while varying $\alpha$ around $\alpha^*$, reveals a parabolic behavior centered around $\Delta e=\Delta e^*\sim 0.01$
corresponding to only $\sim 2.4\%$ difference, independent of cylinder perimeter. 
By interchanging the role of the left and right triangles in the PEPS construction, one can obtain
a degenerate branch with opposite anisotropy. The minima of these parabolas at $\pm \Delta e^*$ therefore define
a pair of degenerate (orthogonal) RVB states, $|\Phi_{\rm L}\big>$ (dressing the left triangles) and $|\Phi_{\rm R}\big>$ (dressing the right triangles),
which transform into each other under inversion symmetry.
Note that the variational energy of this ``Ising doublet", compared in Fig.~\ref{Fig:ener}(b) to the 
DMRG data, is significantly lower than the energy of the NN RVB state\cite{note_brute,note_trimerized}. 

\begin{figure}
\includegraphics[width=0.95\columnwidth]{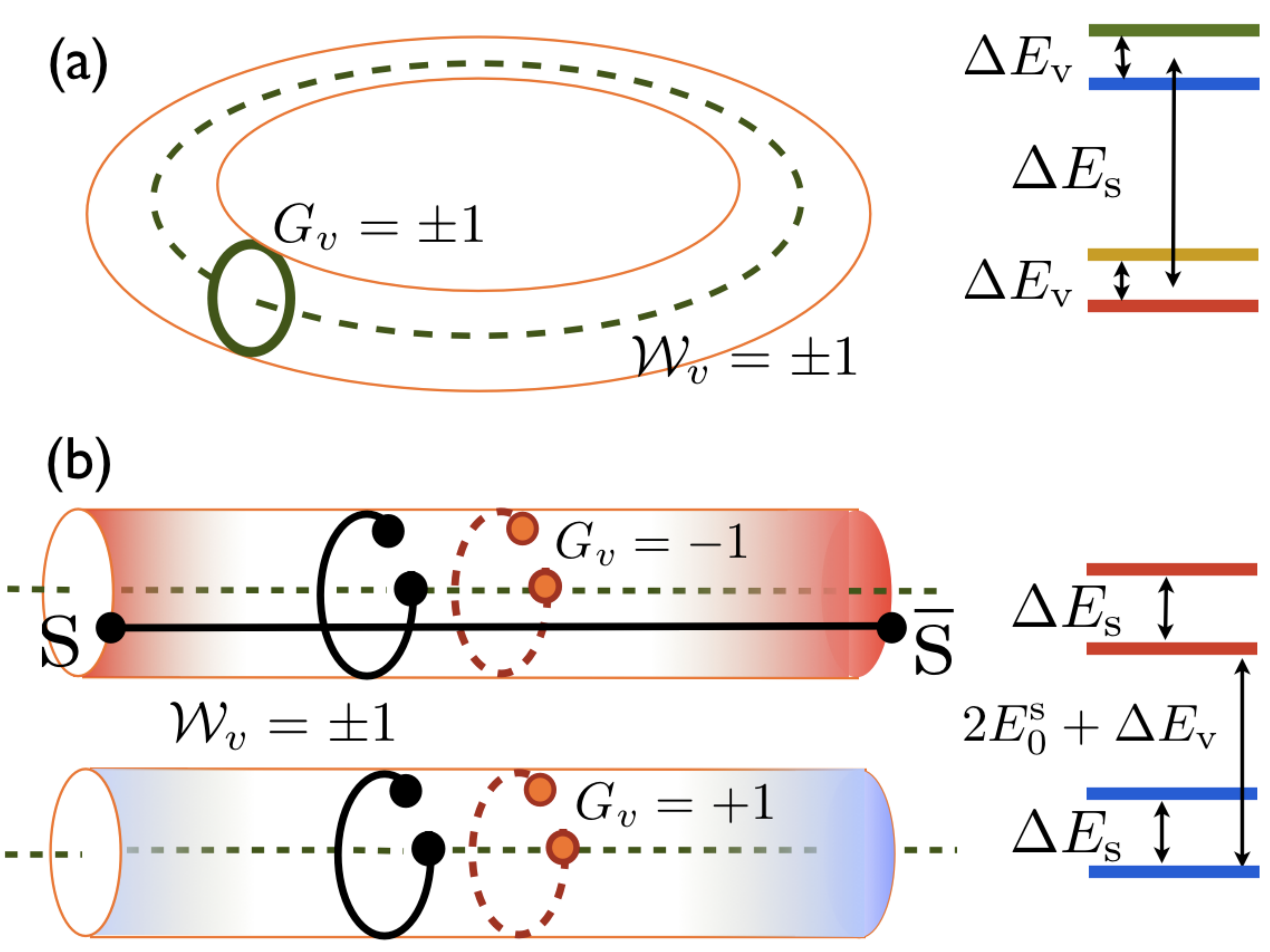}
\caption{The four topological sectors ($G_v=\pm 1$, ${\cal W}_v=\pm 1$) of the $\mathbb{Z}_2$ spin liquids on a thin torus (a) or on a thin cylinder (b). 
Zero or one $\mathbb{Z}_2$-flux can be threaded through the torus/cylinder (dashed green line).
In (b) spinons (visons) are shown by black (orange) dots.
For $G_v=-1$ two spinons s and $\bar{\rm s}$ (with opposite spins) are localized at the cylinder ends. 
For each case, a schematic plot of the low-energy spectrum is shown.  When inversion symmetry is
spontaneously broken, each level becomes doubly degenerate.}
\label{Fig:topo}
\end{figure}

{\it Topological properties --} Let us now consider a {\it finite} thin (horizontal) torus or cylinder,
as depicted in Fig.~\ref{Fig:topo}, of length $N_h$ and circumference $N_v$ such that $N_h\gg N_v$. 
By construction, each component $|\Phi_{\rm L}\big>$ or $|\Phi_{\rm R}\big>$ of the RVB doublet inherits 
the same $\mathbb{Z}_2$ topological properties as the symmetric NN RVB 
spin liquid~\cite{long_paper_topo} or Kitaev's toric code~\cite{kitaev:toriccode}. 
First, the parity $G_v=\pm 1$ of the number of (NN or NNN) singlets cut 
by the vertical (green) closed loop
in Fig.~\ref{Fig:topo}(a), is a conserved quantity and defines two {\it even} and {\it odd} topological sectors. 
In addition, one can consider threading a $\mathbb{Z}_2$ 
flux inside the torus (or the cylinder) along a horizontal loop (dashed line in Fig.~\ref{Fig:topo}(a)). 
Four topological sectors can then be defined by the absence (${\cal W}_v=+1$) or presence (${\cal W}_v=-1$)
of a flux and by the parity $G_v=\pm 1$. 
On the cylinder, changing the parity (fixed by the boundary conditions) is obtained by shifting a line of dimers 
and leaving two extra spinons s and $\bar{\rm s}$ (of opposite spins)
at the two ends. This can be viewed as inserting a "spinon" line joining its two ends (Fig.~\ref{Fig:topo}(b)).
Similarly, a $\mathbb{Z}_2$ flux is added by inserting the
matrix diag(1,1,-1) on all the vertical bonds along a horizontal line, leaving two ``visons" at the cylinder ends.
Although these four sectors cannot be distinguished locally, a spontaneous s${\bar{\rm s}}$ pair (a spontaneous vison pair) whose constituents wind around the torus or the
cylinder in opposite directions before annihilating, as shown in Fig.~\ref{Fig:topo}(b), can detect the presence/absence of a $\mathbb{Z}_2$ 
flux (measure the parity). Such virtual processes lead to splittings $\Delta E_{\rm v}$ and $\Delta E_{\rm s}$  between the (variational) energies in these four topological sectors 
as depicted in Fig.~\ref{Fig:topo}(a,b). Phenomenologically, one expects 
$\Delta E_{\rm v}$ and $\Delta E_{\rm s}$ to be exponentially small in the circumference $N_v$ but to increase linearly with the torus/cylinder length $N_h$, i.e.
\begin{eqnarray}
\Delta E_{\rm s}&=&a N_h N_v \exp{(-N_v/\xi_{\rm spinon})}, \nonumber \\
\Delta E_{\rm v}&=&b N_h N_v \exp{(-N_v/\xi_{\rm vison})}, 
\label{Eq:fits}
\end{eqnarray} 
where $\xi_{\rm spinon}$ and $\xi_{\rm vison}$ are the spinon and vison tunneling lengths.
Note that, in contrast to the torus, the GS energy in the odd parity ($G_v=-1$) sector on the cylinder involves a finite energy cost $2E_0^S$ 
associated to the presence of two spinons at the cylinder ends. Therefore, in the limit 
$N_h,N_v\rightarrow\infty$, at fixed aspect ratio, the RVB liquid is only two-fold degenerate on the cylinder
while it is four-fold degenerate on the torus. 
Note that for the $\alpha=\beta=0$ PEPS, a local parent Hamiltonian has been found
(for which the degeneracy is also equal to 4 on the torus).~\cite{long_paper,note_parent} 

\begin{figure}
\includegraphics[width=0.95\columnwidth]{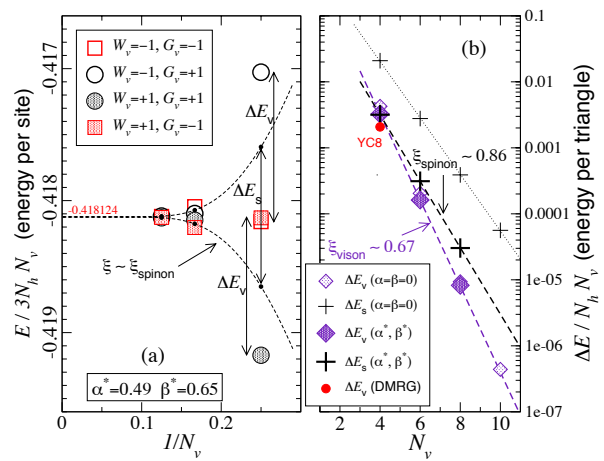}
\caption{(a) Finite size scaling of the RVB energy in the four topological sectors. The energy averaged between even ($G_v=+1$) and odd ($G_v=-1$) boundary conditions (dots) can be fitted as $e_\infty \pm C\exp{(-N_v/\xi)}$
(dashed lines). (b) The splittings in (a) are plotted vs $N_v$ on a logarithmic scale, revealing two different
coherence lengths.}
\label{Fig:splittings}
\end{figure}

Spinon and vison tunneling lengths in the kagome QHAF can be estimated 
using the improved RVB state as a good ansatz.
Finite size scalings of the variational energies (per site)
in the four topological sectors, on the same type of YC cylinders as in DMRG  
and in the limit $N_h\rightarrow\infty$ (we set $E_0^{\rm S}=0$),  are shown in Fig.~\ref{Fig:splittings}(a). 
Results for the splittings shown in Fig.~\ref{Fig:splittings}(b) as a function of the cylinder circumference 
can be accurately fitted according to Eqs.~\ref{Eq:fits}.  The spinon and vison tunneling lengths are found to be quite short, $\xi_{\rm spinon}\simeq 0.86$ and $\xi_{\rm vison}\simeq 0.67$.
It is interesting that these values are quite close to those for $\alpha=\beta=0$, 
$\xi_{\rm spinon}\simeq 1.0$ and $\xi_{\rm vison}\simeq 0.65$, so that we believe that further 
addition of longer range singlet bonds will not significantly modify the coherence lengths.

\begin{figure}
\includegraphics[width=0.46\columnwidth]{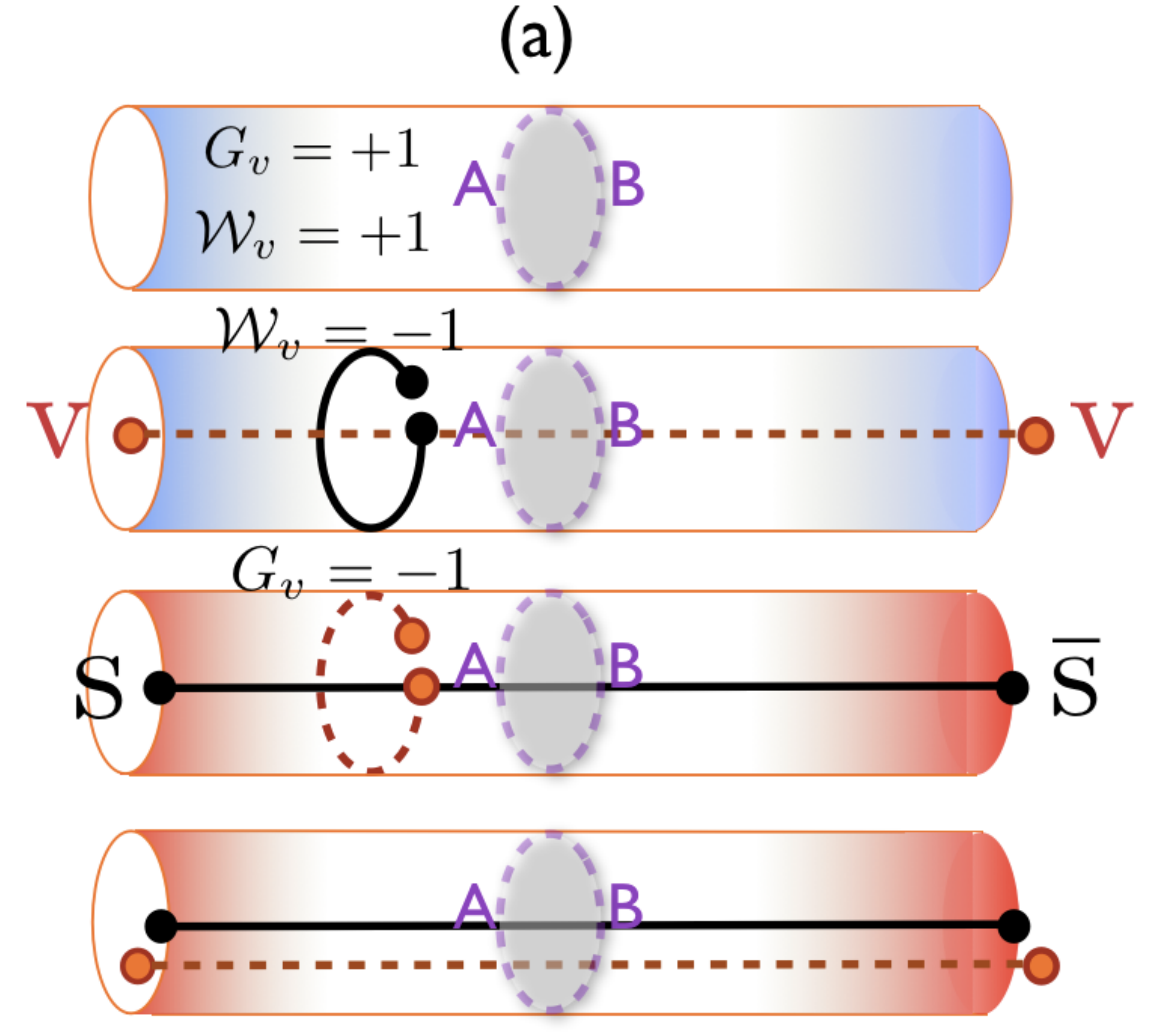}
\includegraphics[width=0.52\columnwidth]{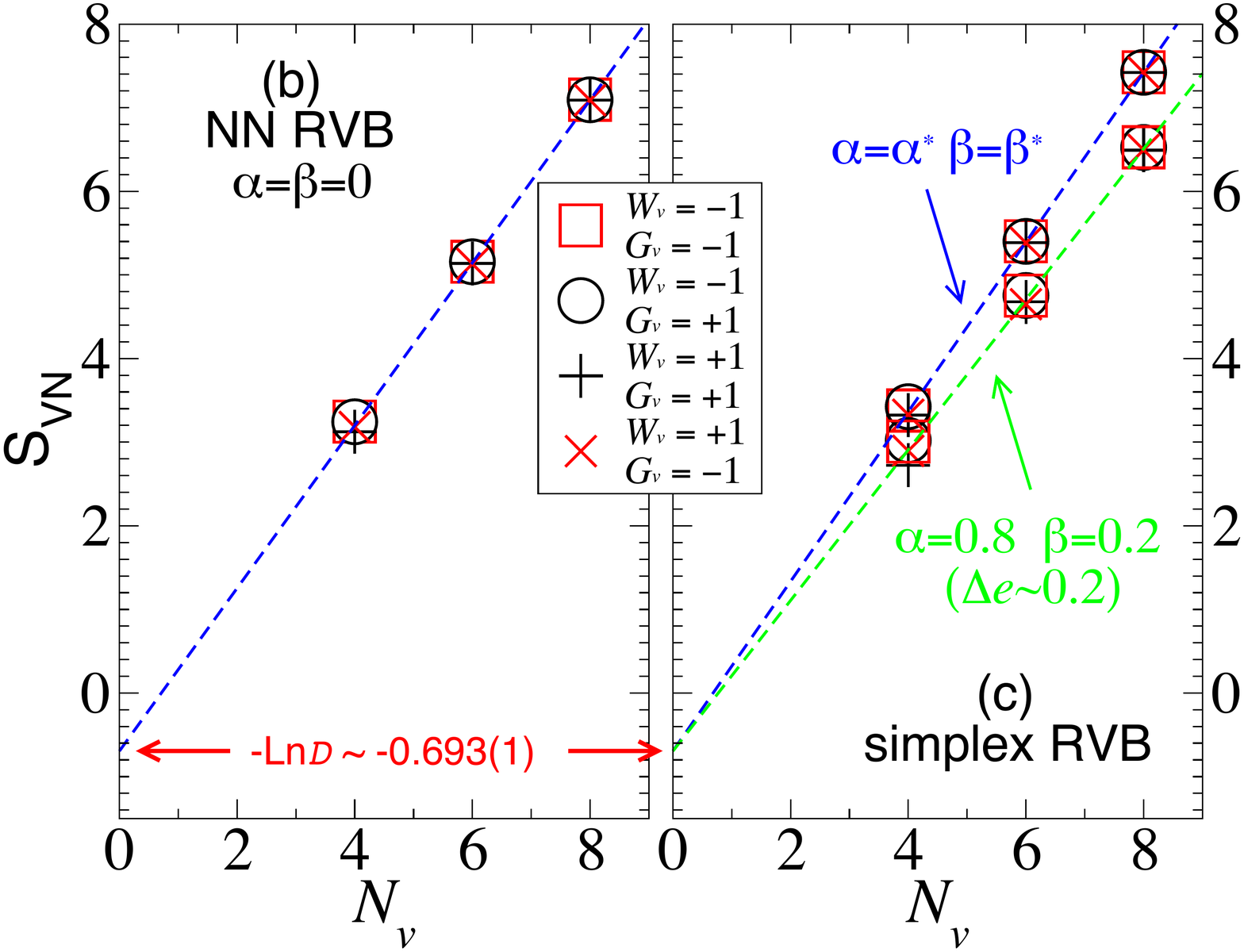}
\caption{(a) Bi-partitions of the long RVB cylinders into A and B semi-cylinders used to compute the EE. By optionally inserting spinon and/or vison lines, four disconnected 
topological sectors can be constructed.
VN entanglement entropy of the symmetric (b) and simplex (c) $\mathbb{Z}_2$ RVB liquids in the four topological sectors as the function of the cylinder perimeter $N_v$. The dashed lines are 
linear fits based on the {\it averages} of the four data points for the two largest (YC12 and YC16) cylinders providing 
$\ln{\cal D}$ with an accuracy of $10^{-4}$.}
\label{Fig:partition}
\end{figure}

{\it Entanglement entropy --} We now turn to the entanglement entropy (EE)~\cite{KP2006} which provides direct 
access to the total quantum dimension $\cal D$.
We consider a bi-partition of the (very long) cylinders in each of the four topological sectors, as shown in Fig.~\ref{Fig:partition}(a). The Von Neumann (VN) EE entropy $-{\rm Tr} \{\rho_A \ln{\rho}_A\}$,
where $\rho_A$ is the reduced density matrix of one semi-cylinder, can be obtained easily in the 
PEPS formalism~\cite{ciracpoilblanc2011,long_paper_topo}.
Results for $\alpha=\beta=0$ and $\alpha\ne 0$, $\beta\ne 0$ are shown in Figs.~\ref{Fig:partition}(b,c). In both cases the data can be fitted according to the {\it area law} $S_{VN}(N_v)=c N_v -\ln{\cal D}$
with ${\cal D}=2$ with high precision, as expected for a $\mathbb{Z}_2$ spin liquid~\cite{EE_RK,EE_TC,zhanggrover}. 
We note that the inclusion of NNN bonds in the improved RVB state
leads to a small increase of the coefficient $c$, i.e. of the wave function entanglement.
We also observe that the difference of the VN EE in the four topological sectors vanishes exponentially
fast with $N_v$. These findings establish clearly the $\mathbb{Z}_2$ topological nature of the NN and 
``simplex" RVB states. 

\begin{figure}
\includegraphics[width=0.7\columnwidth]{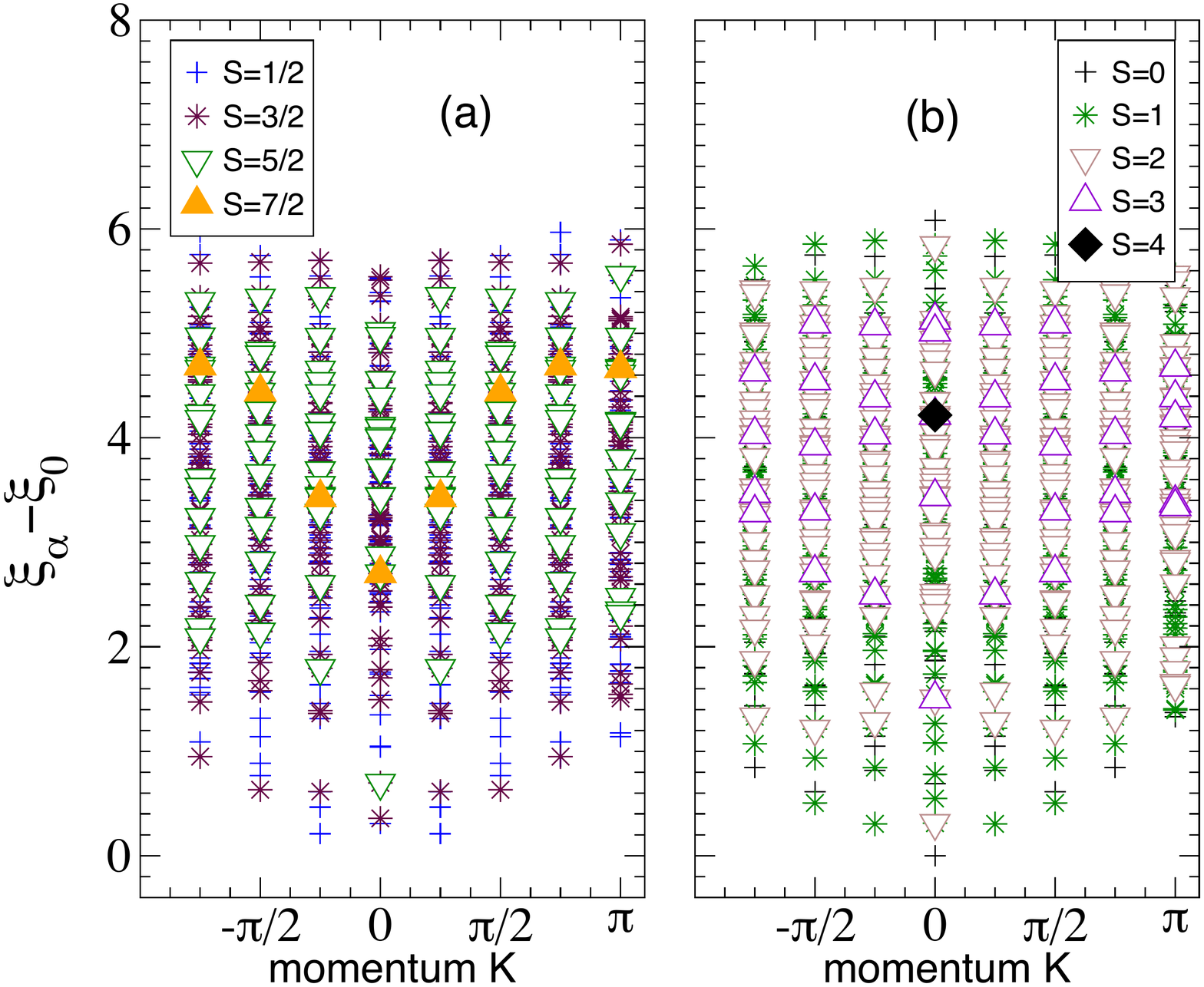}
\includegraphics[width=0.28\columnwidth]{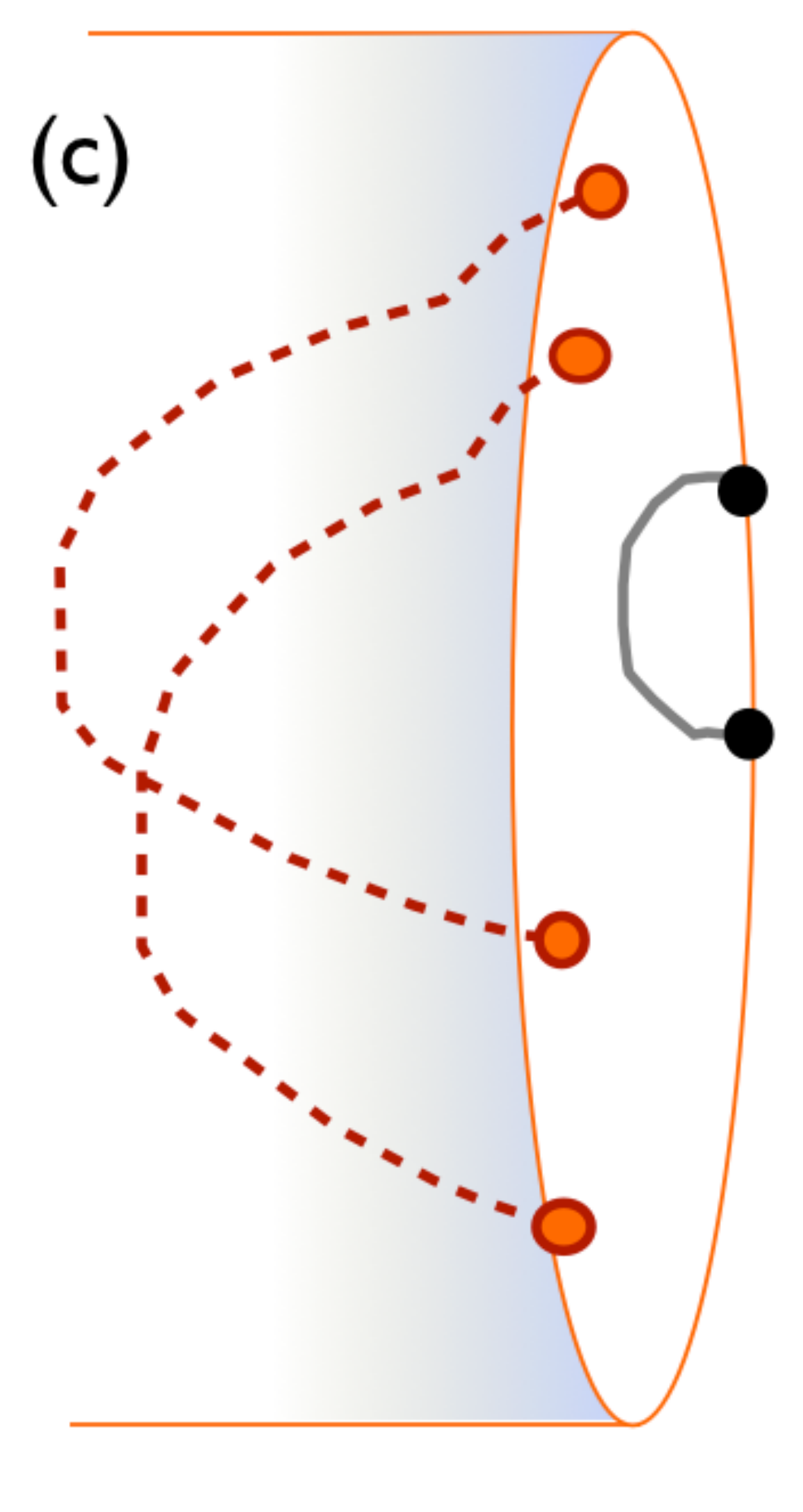}
\caption{Entanglement spectra for a bipartition of a $N_v=8$ (infinite) cylinder
in the $G_v=-1, W_v=+1$ (a) and $G_v=+1, W_v=+1$ topological sectors, for $\alpha=\alpha^*$
and $\beta=\beta^*$. The levels are labelled according to their total spin $S$ (see legends) and momentum 
$K$ along the cut. (c) String representation of the edge states of an infinite semi-cylinder.}
\label{Fig:ES}
\end{figure}

{\it Entanglement spectrum and edge states --}
For fractional quantum Hall states with bulk gap, it has been conjectured that there is 
a deep one-to-one correspondence between the entanglement spectrum~\cite{entspectrum} (ES) of a
partitioned system, i.e. the spectrum of $H_b=-\ln{\rho_A}$, and 
the true edge spectrum of the corresponding $A$ subsystem with a boundary. 
Recently, it has been argued that a similar correspondence applies to {\it non-chiral} gapped quantum 
spin systems~\cite{poilblanc2010}. In fact, the PEPS formalism provides the proof
that $H_b$ acts on emerging degrees of freedom on the edge~\cite{ciracpoilblanc2011}
so that $H_b$ can be viewed as a "boundary Hamiltonian". It is then natural to argue that
such (bond) degrees of freedom become "physical" when the system is "cut" by a real edge
and that their effective interaction is described by the above boundary Hamiltonian. 
We use this conjecture to argue about the existence of gapless edge modes on a RVB cylinder.
The ES of an infinite bi-partitioned cylinder shown in Fig.~\ref{Fig:ES} shows a dense accumulation of levels at
low (pseudo-)energy. A finite size scaling with the cylinder perimeter confirms the gapless nature of the 
ES. Such a feature is robust and was also found for the purely NN RVB state~\cite{long_paper,long_paper_topo}.
The (putative) ES-edge correspondence then implies the existence of gapless edge modes 
for an infinite semi-cylinder with a vertical boundary or for a finite (long) cylinder with two vertical boundaries.
Note that, solvable models with $\mathbb{Z}_2$ topological order, such as the toric code~\cite{kitaev:toriccode},
generically exhibits a gapped ES~\cite{long_paper,schuch2012}.
The existence of gapless edge states~\cite{plaquette_model,cho_2012,edge_z2} in the toric code in fact requires
a particular fine-tuned interaction~\cite{edge_z2}. In the RVB topological phase, the boundary Hamiltonian 
associated to a {\it vertical} edge is an extended $t-J$ model~\cite{long_paper_topo} possessing, 
in addition to topological 
$\mathbb{Z}_2$ invariants, translation and SU(2) symmetries ($\frac{1}{2}\oplus 0$ representation). One then may expect Luttinger liquid-like ($c=1$) 
zero-energy (non-chiral) spin and charge modes~\cite{ogata,note_CFT}.
However, it is not clear whether some degree of fine tuning is realized in the class of effective $t-J$ model 
obtained from $H_b$.
Note that, as in solvable models, the edge states of the RVB liquid
should also have a simple string representation~\cite{edge_z2} as schematically sketched in Fig.~\ref{Fig:ES}(c). 

In summary, we have extended the previous PEPS construction of the NN RVB state for the kagome QAH model
to include fluctuating NNN singlets bonds in the vicinity of defect triangles, resulting in 
a state very close to the optimal wave function  
in the space of translationally invariant PEPS with bond dimension 3.
This novel RVB state dubbed ``simplex spin liquid"
exhibits a weak energy difference ($\sim 2-3\%$) between the two classes of triangles resulting in the loss of the center of inversion, as suggested by recent LED. In addition to $\mathbb{Z}_2$ topological order, Ising ($\mathbb{Z}_2$) order is also present in such a state.
Studying the topological energy splitting
of an infinitely long cylinder as a function of its perimeter gives access to the tunneling lengths of the spinon and vison fundamental fractionalized excitations, which turn out to be of the order of one lattice spacing.  We also argue that 
gapless-energy modes should be present on edges oriented along one of the three crystallographic axes.

DP acknowledges partial supports by the ``Agence Nationale de
la Recherche" under grant No.~ANR~2010~BLANC~0406-0,
the U.S. National Science Foundation under Grant No. NSF PHY11-25915 
(while at Kavli Institute for Theoretical Physics, UC Santa Barbara) and the
CALMIP supercomputer center (Toulouse).
NS acknowledges
support by the Alexander von Humboldt Foundation.
We thank I. Cirac, A.~L\"auchli and G. Misguich for numerous discussions and inputs
and S.~White for providing the DMRG data shown in Figs.~\ref{Fig:ener} and \ref{Fig:splittings}.



\begin{thebibliography}{99}
\bibitem{note_brute}
For $N_v=4$, a brute-force optimization over all $D=3$ tensors gives $e\simeq -0.4222$, only marginally better
than our two-parameter optimal value, $e=-0.4192$.
\bibitem{note_parent}
A parent Hamiltonian can also be constructed for generic $\alpha$ and $\beta$,
although with {\it explicit} inversion symmetry breaking (with the exception of a few particular values).
\end{thebibliography}
\end{document}